# CERCETĂRI PRIVIND STUDIUL VIBRAȚIILOR MECANICE CU SISTEME DE ACHIZIȚII DE DATE

Lenuța SUCIU, Florentina CZIPLE, Cornelia ANGHEL

## RESEARCH ON STUDY MECHANICAL VIBRATIONS WITH DATA ACQUISITION SYSTEMS

The paper presents a new study method of mechanic vibrations with the help of the data acquisition systems. The study of vibrations with the help of data acquisition systems allows the solving of some engineering problems connected to the measurement of some parameters which are difficult to measure having in view the improvement of the technical performances of the industrial equipment or devices.

Cuvinte cheie: vibrații mecanice, achiziții de date, diagrama bloc.

## 1. Introducere

O importanță deosebită este acordată în zilele noastre utilizării calculatorului în prelucrarea numerică a datelor disponibile în urma măsurării unor parametrii fizici ai diferitelor procese industriale şi nu numai. Calculatorul poate fi privit acum ca un sistem de prelucrare numerică constituit pe baza utilizării unui circuit integrat pe scară largă de tip microprocesor, microcontroler sau procesor numeric de semnal. În majoritatea aplicațiilor de acest tip se pune problema de a obține informaţii despre aceste procese fizice în vederea memorării şi redării pentru comunicaţie sau pentru control. Un astfel de proces este caracterizat prin mărimi fizice care ulterior pot fi transformate în semnale electrice analogice utilizând traductoare. Prelucrarea acestor semnale se poate face utilizând tehnici analogice sau numerice. În vederea unei prelucrări numerice este necesară transformarea semnalelor analogice în semnale numerice cu un sistem de achiziţie date.

## 2. Funcţiile sistemelor de achiziţie şi prelucrare a datelor

Sistemele de achiziţie şi prelucrare a datelor sânt sisteme de complexitate variată, realizate în scopul:
- urmăririi, monitorizării unor fenomene sau procese;
- măsurării unei mulţimi de mărimi în procesul de experimentare a modelelor

funcţionale ale unor produse, sisteme;
- testării produselor finite;
- observării şi controlului proceselor de producţie;
- monitorizării şi controlului traficului terestru, naval şi aerian.

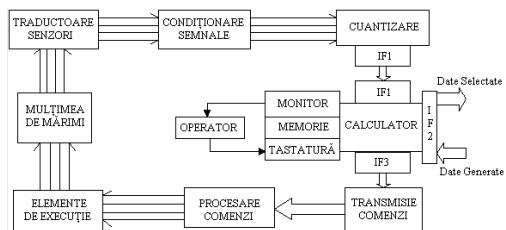

Fig. 1 Structura generală a SAPD cu un calculator.

În cadrul unui SAPD, mediul de programare are rol de a prelua şi prelucra informaţia primită de la digitizor (placă de achiziţie), adică îndeplineşte următoarele funcţii: controlează şi comandă întregul sistem de achiziţie, preia valorile achiziţionate digitizate printr-o magistrală de comunicaţie, reconstituie semnalul din eşantioane (conversie digital-analogică), determină anumite valori caracteristice ale semnalului (amplitudine (valoare de vârf) etc.), realizează o filtrare soft a semnalului, realizează o analiză Fourier (spectrală), efectuează diverse operaţii aritmetice (adunare, înmulţire, radical, integrare, derivare etc.).

Mediul LabVIEW este de fapt un limbaj grafic de programare la care codul nu este scris sub formă de text, ci cu ajutorul unor pictograme. Programele (aplicațiile) realizate cu LabVIEW se numesc instrumente virtuale deoarece reproduc funcționarea instrumentelor reale cum ar fi: ampermetre, voltmetre, osciloscoape, ohmetre, multimetre etc.

La un instrument de măsurare obișnuit se disting două părți:
a) Interfața cu utilizatorul (afișajul, butoane pentru stabilirea domeniilor) căreia îi corespunde, la aplicațiile realizate cu LabVIEW, fereastra numită *Panoul Frontal*;
b) Mecanismul de funcționare pe baza căruia funcționează instrumentul, căruia îi corespunde, la aplicațiile realizate cu LabVIEW, fereastra numită *Diagrama Bloc*.

### 3. Utilizarea practică a mediului de dezvoltare LabVIEW în studiul vibrațiilor unei lamele elastice

Vibrațiile sunt fenomene dinamice, care apar în medii elastice în urma unei excitații locale și care se propagă în interiorul mediului sub forma unor oscilații. Vibrația este o oscilație mecanică în jurul unui punct de referință și definește mișcarea unui sistem mecanic. Mediul trebuie să fie suficient de mare pentru a se putea vorbi de o excitație locală, respectiv ca acesta să se propage prin oscilații. Vibrația se caracterizează prin amplitudine, viteză, accelerație și spectru de frecvențe. Vibrația este de multe ori distructivă, pe de altă parte este latura deranjantă a unui lucru folositor, dar poate fi generată și în mod intenționat pentru realizarea unor cerințe.Mărimile care caracterizează mișcarea (vibrația) sistemului, adică deplasarea, viteza, respective accelerația, se definesc conform relațiilor:

$$d = D \sin \omega t \quad \text{deplasarea}$$
$$v = dd/dt = D\omega \cos \omega t \quad \text{viteza}$$
$$a = d^2 d / dt^2 = D\omega^2 \sin \omega t \quad \text{accelerația}$$

În realitate aceste mărimi sunt mai complexe legea de variație nu este sinusoidală. De aceea se face o înregistrare a vibrației, se descompune în componente care deja sunt sinusoidale, se face o analiză spectrală și pe baza acesteia se determină natura vibrației. Cunoașterea mărimilor caracteristice vibrațiilor are o mare importanță în tehnică. Amplitudinea vibrațiilor informează despre jocurile existente între piese, accelerația vibrației informează despre intensitatea forțelor de solicitare care acționează din cauza vibrației, iar viteza informează despre zgomotul acustic produs de mediul care vibrează, dar şi despre energia vibrației.

### 4. Prezentarea sistemului de achiziție și prelucrare a vibrațiilor

Pentru măsurarea vibrațiilor se utilizează un accelerometru triaxial ceramic 8762A5T. Accelerometrul este fixat pe o lamelă elastică, la un capăt, iar celălalt capăt al lamelei este încastrat, lamela fiind supusă unei forțe $F$, aplicată manual. Ca și la majoritatea accelerometrelor sensibilitatea este dată de raportul ieșirii electrice de a aplica accelerații, la ieșire obținându-se o tensiune de impedanță scăzută, care este proporțională cu accelerația aplicată. Datorită impedanței scăzute nu se necesită conexiuni speciale și transmiterea la distanță este posibil să fie realizată cu minimum de zgomot.

Pentru sesizarea vibrațiilor, accelerometrul fixat pe lamela elastică convertește accelerația într-o mărime electrică, care este proporțională cu forța aplicată pe elementul ceramic intern (piezoelectric), variabila mecanică (accelerația) fiind obținută dintr-o măsurare a forței.

Ansamblul este compus dintr-o bară centrală, un element ceramic piezoelectric, o masă seismică și o săgeată de preîncărcare. În momentul funcționării unitatea transmite o mișcare perpendiculară către bază. Când accelerometrul este atașat la o structură vibrantă, masa seismică exercită o forță pe elementul ceramic piezoelectric. Această forță aplicată determină materialul piezoelectric să producă o mărime electrică. Forța este egală cu masa ori accelerația (*legea a doua a lui Newton F = ma* ), rezultatul obținut este proporțional cu accelerația, atâta timp cât masa *m* este constantă.

Componentele sistemului de achiziție și prelucrare a vibrațiilor lamelei elastice sunt date de modulele SCXI 1530 și SCXI 1600, circuitul de condiționare al semnalului (vibrației), care este chiar placa accelerometrului utilizat, respectiv placa de achiziție propriu-zisă, ambele fiind alimentate prin intermediul sursei de alimentare SCXI 1000.

În figura 2 este prezentată diagrama bloc a instrumentului virtual utilizat în prelucrarea accelerației.

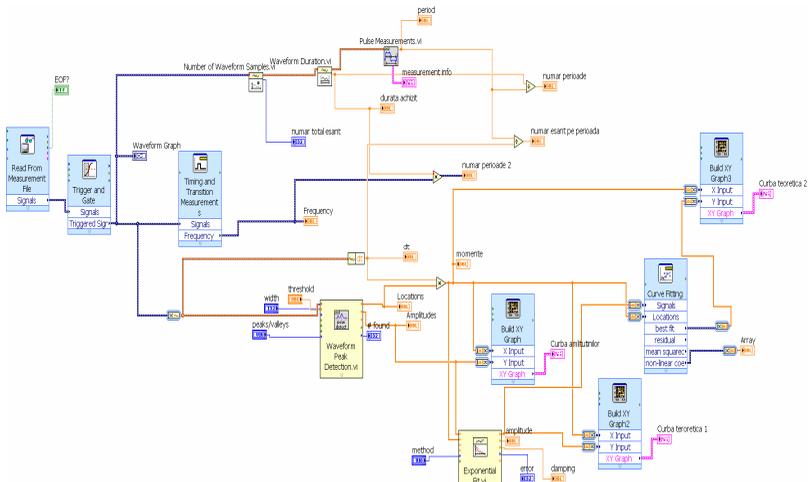

Fig.2 Diagrama bloc a instrumentului virtual utilizat în prelucrarea accelerației.

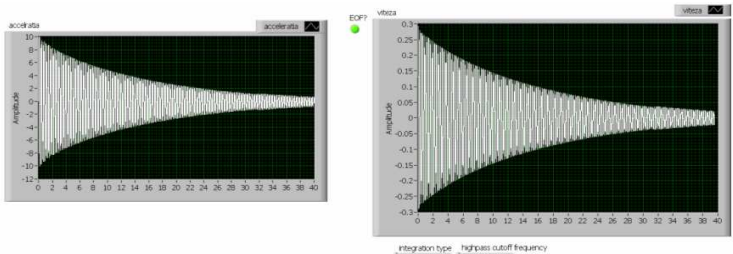

Fig.4 Panoul frontal al instrumentului virtual utilizat în prelucrarea accelerației și a vitezei.

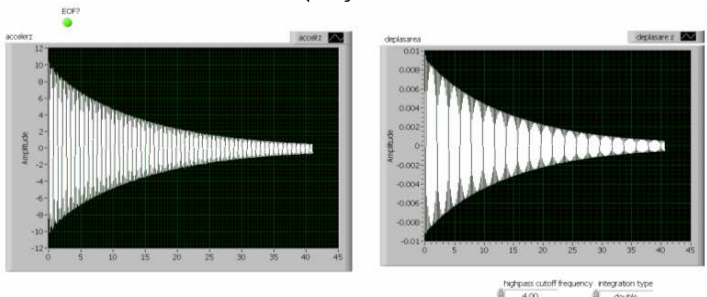

Fig.5 Panoul frontal al instrumentului virtual utilizat în prelucrarea accelerației și a deplasării.

## 5. Concluzii

Studiul vibrațiilor permite inginerului să rezolve o serie de probleme ce se întâlnesc în tehnica:
- ➢ stabilirea turațiilor critice ale arborilor mașinilor;
- ➢ proiectarea suspensiilor autovehiculelor şi vagoanelor;
- ➢ proiectarea izolării antivibratoare a mașinilor staționare;
- ➢ măsurarea vibrațiilor, la cele mai variate maşini si construcții şi aprecierea efectului lor asupra oamenilor, mașinilor şi clădirilor;
- ➢ proiectarea mașinilor vibratoare cu diferite utilizări, încercări la oboseală, compactări de terenuri, operații de transport, sortat, etc.;

Ş.l.dr.ing. Lenuța SUCIU,
Universitatea ,, Eftimie Murgu" Reşița, membru AGIR
e-mail: l.suciu@uem.ro
Conf.dr.ing. Florentina CZIPLE,
Universitatea ,, Eftimie Murgu" Reşița, membru AGIR
e-mail: f.cziple@uem.ro
Ş.l.dr.ing. Cornelia ANGHEL,
Universitatea ,, Eftimie Murgu" Reşița, membru AGIR
e-mail: c.anghel@uem.ro